# 北京师范大学

# 本科生毕业论文(设计)

毕业论文（设计）题目：

## 量子计算机可实现的阻挫量子相变的研究

部　院　系：　物理学系

专　　　业：　物理学（基地）

学　　　号：　201911061135

学 生 姓 名：　陈自立

指 导 教 师：　邵　慧

指导教师职称：　教　授

指导教师单位：　北京师范大学物理学系

2024 年 5 月 27 日

## 北京师范大学本科生毕业论文（设计）诚信承诺书

本人郑重声明： 所呈交的毕业论文（设计），是本人在导师的指导下，独立进行研究工作所取得的成果。除文中已经注明引用的内容外，本论文不含任何其他个人或集体已经发表或撰写过的作品成果。对本文的研究做出重要贡献的个人和集体，均已在文中以明确方式标明。本人完全意识到本声明的法律结果由本人承担。

本人签名： 年 月 日

## 北京师范大学本科生毕业论文（设计）使用授权书

本人完全了解北京师范大学有关收集、保留和使用毕业论文（设计）的规定，即：学校有权保留并向国家有关部门或机构送交论文的复印件和电子版，允许毕业论文（设计）被查阅和借阅；学校可以公布毕业论文（设计）的全部或部分内容，可以采用影印、缩印或扫描等复制手段保存、汇编毕业论文（设计）。保密的毕业论文（设计）在解密后遵守此规定。

本论文（是、否）保密论文。
保密论文在__*__年__*__月解密后适用本授权书。

本人签名： 年 月 日
–
–
导师签字： 年 月 日

# 量子计算机可实现的阻挫量子相变的研究

## 摘 要

量子计算机具有并行计算、纠缠效应的特点，在密码分析、大数据处理等方面具有独特的优势。然而这类计算机尚未发展成熟，许多性能有待考量，需传统计算机数据作参考，尤其是量子相变的模拟。可以选择二维阻挫晶格体系作为量子相变的研究对象。目前，阻挫正方晶格和阻挫三角晶格的研究在传统计算机上已取得显著进展，而关于六角蜂窝晶格的研究较少。

论文分为四个部分。第一部分描述了量子计算机以及量子相变和序参量的相关背景；第二部分解释了量子蒙特卡洛算法的基本思路；第三部分进行数值模拟，在低温条件下模拟了不同横向磁场对序参量的影响，展示了不同尺寸晶格的结果；第四部分是总结展望，讨论了磁场范围的影响，并与正方晶格的结果相比较。

**关键词**：伊辛模型，阻挫，序参量，量子蒙特卡洛算法，六角蜂窝晶格

# Study on Frustrated Quantum Phase Transition Achievable by Quantum Computing

Abstract

Quantum computers, with parallel computing and entanglement effects, excel in cryptography analysis and big data processing. However, they are not fully developed yet, and their performance needs further evaluation. Traditional computer data, especially in simulating quantum phase transitions, are still needed for reference. Two-dimensional frustrated lattice systems can be chosen for studying quantum phase transitions. Currently, significant progress has been made in the study of frustrated square and triangular lattices using traditional computers, while research on hexagonal lattices is limited.

This paper consists of four parts. The first part introduces the background of quantum computers and the concept of quantum phase transitions, with the selection of order parameters in hexagonal lattices. The second part elaborates the ideas of the quantum Monte Carlo algorithm. The third part presents numerical simulations, exploring the impact of different transverse magnetic fields on order parameters under low-temperature conditions and showcasing results for various lattice sizes. The fourth part summarizes and looks ahead, comparing the results with those of square and triangular lattices as well as relevant theoretical analyses.

# 目录

# 第1章  引言

## 1.1 量子计算机

现如今，传统计算机发展势头逐渐变缓，在算力优化上面临着难以解决的问题。想要提高算力，就需要缩小晶体管的体积、提高晶体管的集成密度。而晶体管的体积已快接近物理极限，进一步缩小体积将产生显著的量子效应干扰晶体管的正常工作。可见，传统计算机的很难取得突破了。如今，众多领域都需要处理海量数据，例如加密破解、气象预报、人工智能、药物设计等，传统计算机的算力显得有所不足。相较而言，量子计算机正在快速发展，在需要处理大量数据的领域中表现出远优于前者的性能。自1980年保罗.本尼奥夫提出量子计算机的概念，再到1994-1996年量子算法取得突破，随后的二十多年来，各国政府、国际大公司、科研机构都为量子计算投入了大量人力物力，并产出了相应成果，例如，D-Wave公司现已推出基于伊辛模型的量子退火计算机，IBM公司已经实现量子通用计算机的商业化，我国的"九章"玻色量子计算机问世，快速解决了高斯玻色取样问题[1]。

量子计算机的优越性能主要源于它的并行计算。传统计算机依赖经典物理比特，通过晶体管的导通、断开状态实现0和1，其比特位状态确定，只能存在导通或断开中的一种。量子计算机依赖的是量子物理比特，后者可叠加多种物理态，例如，低温超导环能实现上自旋态与下自旋态的叠加，比特位同时映射为0和1。经典和量子比特具有截然不同的性能。以对应着16种不同的状态的4比特为例，4个经典比特每次只能表示一种状态，单次运算对单状态进行操作；一个量子比特可同时处于两种模式的叠加，每多耦合一个量子比特，整体的状态数目翻倍，4量子比特相干叠加后就可同时表征出16种状态，单次运算就等效应同时对其中的16种状态进行操作。可见，传统计算机每次只处理单个目标，而量子计算机每次能同时处理多个目标，从而实现并行运算，计算效率优势斐然。

## 1.2 量子相变

### 1.2.1 相变中的量子效应与阻挫现象

目前，量子计算机在多个领域都展现出了巨大的潜力，尤其是模拟量子相变[2]。相变指系统在外部参数的变化下，某种宏观属性发生变化的过程。经典相变依赖热涨落，改变温度，物体将会跨越不同的相。量子相变有别于经典相变，它依赖量子涨落使系统在大量状态之间跃迁，改变外加磁场等参数可以影响量子涨落，进而使系统跨越不同的相[3]。

零温下，系统随哈密顿量参数变化而产生的相变即量子相变，跨越临界点，系统基态将发生变化。一种典型的哈密顿量如下所示，g表示无量纲耦合量，此时系统的基态能解随着g的变化可能出现非解析点（能级交叉的点或者避免能级交叉的极限情况），这样的

点即量子相变点。

$$H(g) = H_0 + gH_1 \tag{1-1}$$

下图 1 是量子相变的示意图[3]，由温度T与参数g描述——本文研究的g是横向磁场。在到达零温前，相变临界线附近的阴影区域主要由经典热涨落规律支配；接近零温的过程，该区域缩小，量子效应逐渐占据优势；完全到达零温时，哈密顿量参数为临界值，$g = g_c$，对应着量子临界点。然而，根据热力学第三定律，零温的量子相变不能通过实验验证，因此对量子相变的研究主要依赖理论求解、低温实验、数值模拟极低温度下的行为等方式。

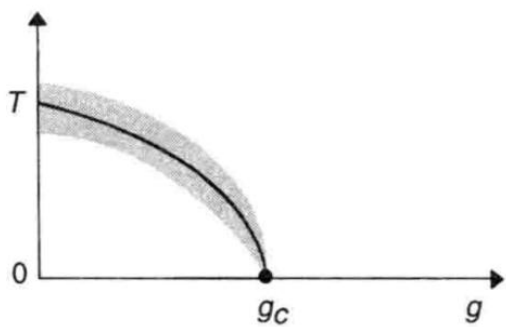

图 1 典型的量子相变示意图

本文关注的是伊辛模型下的二维六角蜂窝晶格的量子相变，体系依靠几何阻挫产生强烈的量子效应，进而带来丰富的量子相变。阻挫是自然界普遍存在的现象，表现为体系内各个单元无法同时处于各自能量最低的状态，导致体系存在大量相互竞争的基态。其中，几何阻挫是一种典型的类别，它源于晶格格点之间的磁耦合能在空间上的特殊分布。下图展示了一个简单的阻挫单元。三个格点在z方向具有两种本征自旋值，上自旋$s = 1/2$ 和下自旋$s = -1/2$，任意两个格点i、j之间的耦合势能有：$e_{ij} = J_{ij} s_i s_j$。其中$s_i, s_j$分别代表相邻格点自旋值，$J_{ij}$是磁耦合系数，$J_{ij}<0$ 是铁磁耦合，$J_{ij}>0$ 是反铁磁耦合。例如上图，左下角格点是下自旋态，中间的格点是上自旋态，此时不论右下角格点为何种自旋，都无法满足所有的耦合势能同时处于最低值，始终存在 $e_{ij}>0$ 的阻挫键——这样的三角单元就有六种不同的基态。

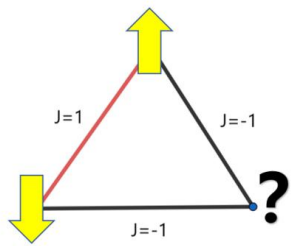

图 2 具有阻挫的三角模型

## 1.2.3 序参量的介绍和引入

关于几何阻挫的量子相变有很多研究，而这些研究中的量子相变通常用序参量来描述。序参量反应了系统的某种对称性，它应该这样构造：在对称性较高的系统（无序相）为零，在对称性低的系统（有序相）非零[4]。在铁磁相变中，序参量即为磁化强度：系统处于顺磁态时，磁矩单元杂乱无章地排布，系统的磁化强度为零，经旋转变换后磁化强度不变，系统的对称性较高；系统处于铁磁态时，磁矩单元倾向于有序排列，磁化强度具有方向性，其值不为零，系统的对称性较低。在数值模拟中，磁化强度可写为如下形式，其中N是系统的格点数，S是格点i上的磁矩。

$$m = \frac{1}{N} \sum_i S_i \tag{1-2}$$

反铁磁系统的序参量$m_{anti}$则基于子晶格的磁化强度$m_1$和$m_2$进行构建，如下所示。如果系统的格点磁矩无序排列，$m_1$、$m_2$都为零，则序参量$m_{anti}$为零；如果系统格点的磁矩交错有序排列，$m_1$和$m_2$分别达到"1"和"-1"，则系统的$m_{anti}$绝对值达到最大。

$$m_{anti} = \frac{1}{2}(m_1 - m_2) \tag{1-3}$$

上述铁磁和反铁磁系统的基态位形数少，不具备几何阻挫。要构造几何阻挫的晶格体系，可以根据Villain原则[5]在晶格的多边形耦合链中插入奇数个反铁磁耦合作用（$J_{ij}>0$）。此时系统的序参量有所不同，下面先考察三角晶格和正方晶格序参量的构造。

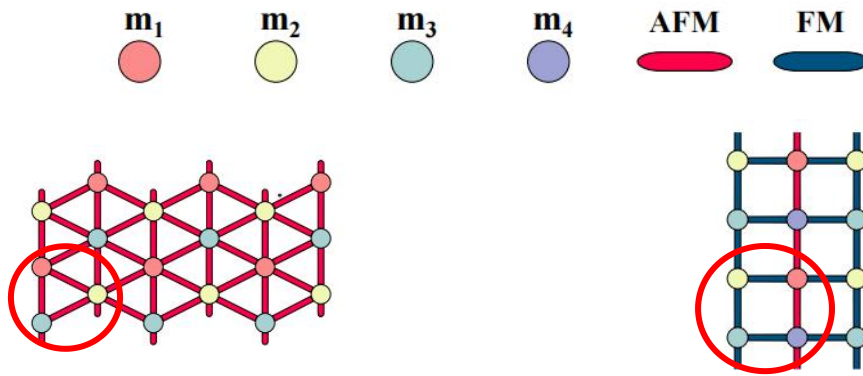

图 3.1 三角晶格阻挫单元[2]　　　　　图 3.2 正方晶格的阻挫单元[2]

上图 3.1[2]是满足Villain原则的三角晶格的示意图，三种颜色的格点对应三种布拉维子晶格，红色的边表示反铁磁作用。序参量$m_{tri}$如下(1-4)式，子晶格磁化强度为$m_s$，按照一定的相位角$e^{i\phi}$对序参量$m_{tri}$做贡献。当每个三角形阻挫单元（红色圈）处于相同的阻挫键数量最少的自旋位形时，整个系统也处于基态（伊辛模型磁耦合能描述的基态）。$m_{tri}$对于这 6 种基态得到的模长是相等，相位角依次有：$e^{n\pi i/3}, n = 1,2,...,6$。$m_{tri}$的模长越大，说明三角单元自旋的形式越单一，自旋排布趋于有序，系统越接近基态；如果系统处

于非基态的情况，三角阻挫单元内的$m_s$往往处于无序排布的状态，此时$m_{tri}$的模长趋于零。

$$m_{tri} = \frac{1}{\sqrt{3}}(m_1 + e^{2\pi i/3} m_2 + e^{4\pi i/3} m_3) \tag{1-4}$$

$$m_s = \frac{3}{N}\sum_{i \in s} \sigma_i^z \tag{1-5}$$

上图 3.2[2]是满足 Villain 原则的正方晶格的示意图，不同颜色的格点对应四种子晶格，蓝色边表示铁磁作用，红色边表示反铁磁作用，红色圈代表一个正方形阻挫单元。正方形晶格的初级序参量$m_{square}$如下（即下式的 m）。同样的，当每个正方形阻挫单元处于相同的阻挫键数量最少的自旋位形时（记为单元基态），整个系统也处于基态（伊辛模型磁耦合能描述的基态）。$m_{square}$对于这 8 种基态具有相等的模长，相位角依次有：$e^{n\pi i/4}, n = 1,2,\dots,8$。八种单元基态如下图 4.1 所示，系统处于平衡状态时典型的$m_{square}$分布情况如下图 4.2 所示[6]。

$$m_{square} = \frac{1}{2}(e^{\pi i/8} m_1 + e^{3\pi i/8} m_2 + e^{5\pi i/8} m_3 + e^{7\pi i/8} m_4) \tag{1-6}$$

$$m_s = \frac{4}{N}\sum_{i \in s} \sigma_i^z \tag{1-7}$$

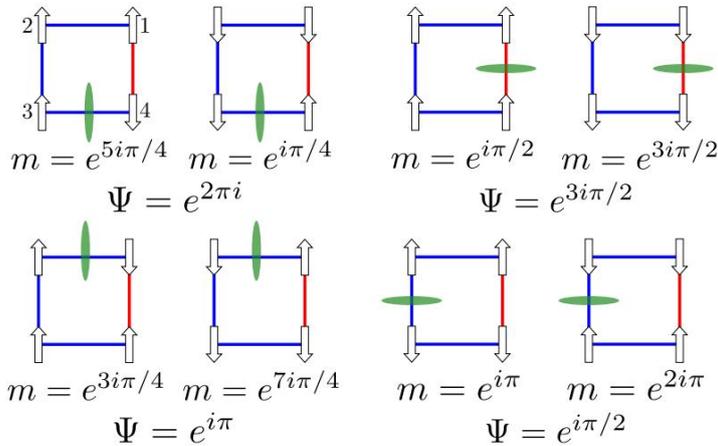

图 4.1 正方形阻挫单元的八种基态[6]

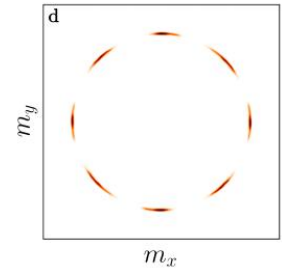

图 4.2 初级序参量[6]

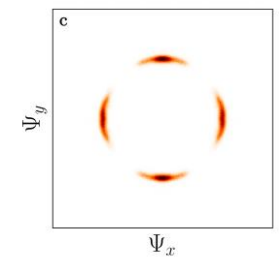

图 4.3 次级序参量[6]

此外，研究者还引入了次级序参量$\psi_{square}$（即上图的 ψ）描述正方单元内二聚体（即相邻格点耦合能大于零的阻挫键）的出现频率。基于二聚体密度公式$d_{i,\alpha} = \frac{1}{2}(1 + J_{ik}/J \cdot \sigma_i^z \sigma_{k,\alpha}^z)$，可以判断近邻的 i、k 格点间关于沿 α 方向是否存在二聚体。经傅里叶变换后：$\tilde{d}_\alpha(\mathbf{q}) = \frac{1}{N}\sum_i e^{i\mathbf{q}\cdot\mathbf{r}_i} d_{i,\alpha}$，合成次级序参量$\psi_{square}$如下：

$$\psi_{square} = 2\tilde{d}_x(0,\pi) + 2\tilde{d}_y(\pi,0) \tag{1-8}$$

当每个正方形阻挫单元只含有一种类型的阻挫键时，则$\psi_{square} = e^{n\pi i/2}$，$n = 1,2,3,4$，对应上图 4.3。

可见，当所有正方形阻挫单元处于相同的单元基态时，整个系统也处于基态，相应的序参量模长最大、具有明显的方位性。随着温度的升高，系统的自旋位形逐渐趋于无序，序参量图像逐渐收缩，模长减小，如图 5.1 所示（以次级序参量为例）。此外，还有研究者对比[7]不同尺寸下的正方晶格的次级序参量（T=1.5），研究发现系统尺寸越大，分布图像越来越集中于四个特定的方位，如下图 5.2 所示。

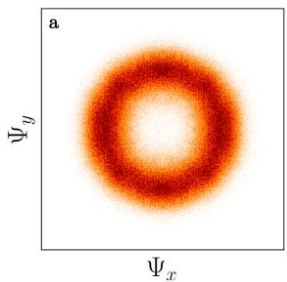
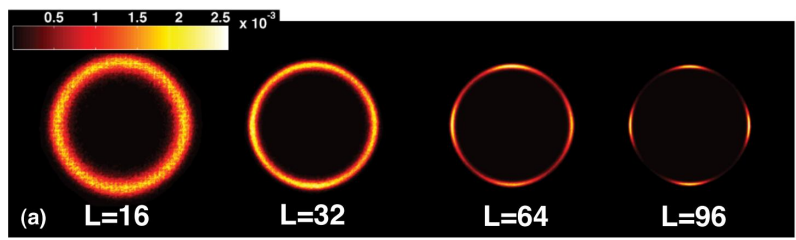

图 5.1 模长渐趋于零[6]　　　　图 5.2 尺寸越大，序参量分布越集中[7]

本文关注的是几何阻挫的六角蜂窝晶格。Kitaev 曾提出了一种化学键各向异性的六角蜂窝结构，能产生本征阻挫[8][9]；也有研究者整理了属于具有本征阻挫的六角蜂窝结构的材料，测量了比热、导热率等参数[10]。但是，关于几何阻挫的六角蜂窝及其序参量的研究较为少。作为对这方面的进一步探索，本文设置了一种符合Villain原则的六角蜂窝晶格（图 6.1），并结合三角晶格、正方晶格序参量的形式，做尝试性推广。

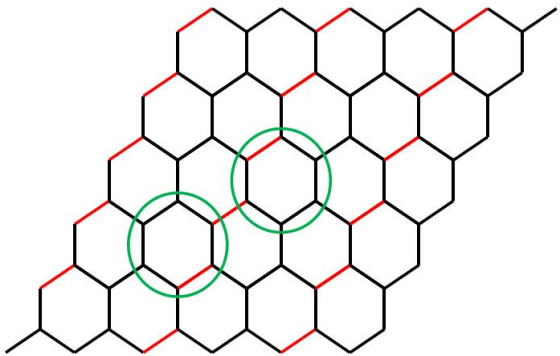

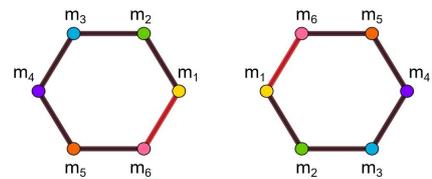

图 6.2 六边形阻挫单元

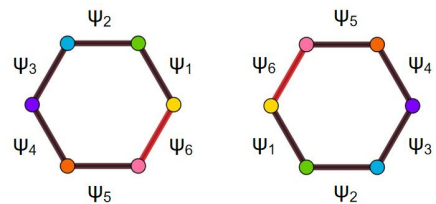

图 6.1 几何阻挫的六角蜂窝晶格　　　　图 6.3 阻挫键类型

上面两种晶格的序参量定义方式和系统的基态有关，当系统处于基态时，其自旋位形处于伊辛耦合能描述下的某种有序状态，对应的序参量模长最大且具备方位性。本文从基态与序参量的对应关系出发，探索六角蜂窝晶格的情况。基态意味着每个多边形单元平均分配到的单元阻挫键数量最少，因此可以从六边形阻挫单元阻挫键数最少的单元基态出发，

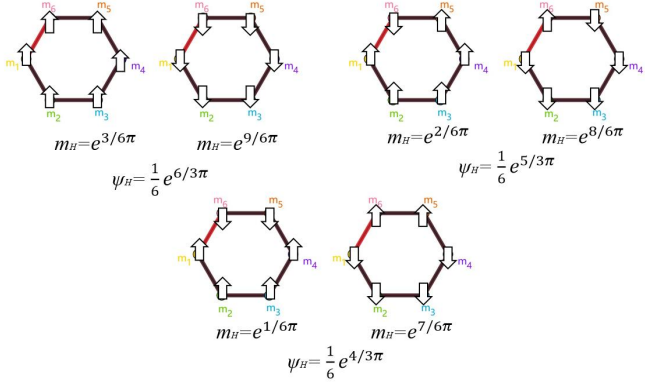

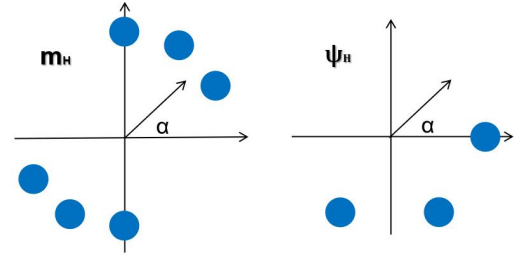

图 7.1 带来基态的六种阻挫单元　　　　　图 7.2 相应的序参量示意图

找到系统的部分基态位形，进而确定序参量的形式。如上图 6.1 所示，绿色圈是本文考察的六边形单元对象，系统的格点都能被这两种情况包含。六边形单元内的格点编号如图 6.2 所示，图 6.3 是相应的阻挫键类型编号。这样编号的格点，在蜂窝晶格上仍具有周期性，可以将相同编号格点（例如图 6.1 绿色圈内的两个相同编号的格点）两两一组看成一个新格点，不同编号描述的新格点分属于六种不同的子晶格。本文发现，当系统的六边形阻挫单元都处于图 7.1 所示的六种位形时，系统整体便处于基态。为了实现这类基态与序参量的相关性，仿照正方晶格序参量的相位因子，尝试写出六角蜂窝晶格的初级序参量如下：

$$m_H = \frac{1}{a}(e^{\frac{\pi i}{12}}m_1 + e^{\frac{3\pi i}{12}}m_2 + e^{\frac{5\pi i}{12}}m_3 + e^{\frac{7\pi i}{12}}m_4 + e^{\frac{9\pi i}{12}}m_5 + e^{\frac{11\pi i}{12}}m_6) \qquad (1-9)$$

$$m_s = \frac{1}{N}\sum_{i\in s}\sigma_i^z \qquad (1-10)$$

上式 $a = \sqrt{2} + \sqrt{6}$，N 为系统格点数。当每个六边形阻挫单元处于图 7.1 所示位形时，系统的 $m_H$ 对于这六种基态具有相等的模长，有 $m_H = e^{n\pi i/6}$，n = 1，3，5，7，9，11，12，如图 7.2 所示。对于一般的图，如果阻挫单元大量地处于图 7.2 所示的某种类型时，则 $m_H$ 的模长较大，且具有 6 种方位性；如果系统自旋处于无序状态，由于平均效应，模长趋于零，序参量 $m_H$ 的分布向中心收缩，不再具有方向性——类似上图 5 代表的变化趋势。

此外，参照正方晶格，可知次级序参量能描述阻挫单元存在的阻挫键位置类型。从这一点出发，引入次级序参量的形式：

$$\psi_H = e^{\frac{\pi i}{3}}\psi_1 + e^{\frac{2\pi i}{3}}\psi_2 + e^{\frac{3\pi i}{3}}\psi_3 + e^{\frac{4\pi i}{3}}\psi_4 + e^{\frac{5\pi i}{3}}\psi_5 + e^{\frac{6\pi i}{3}}\psi_6 \qquad (1-11)$$

如上图 6.3 所示，$\psi_k$ 代表六边形阻挫单元上耦合边的阻挫键情况，如果耦合边上存在阻挫

键，则$\psi_k$取为1/N，如果不存在则取为0。当系统的六边形阻挫单元都处于图7.1所示的某种类型时，次级序参量有：$\psi_H = 1/6 \cdot e^{n\pi i/3}$，n = 4，5，6，如上图7.2所示。对于一般的$\psi_H$图，如果系统的某种阻挫键类型占优势，则$\psi_H$模长较大，且具有方向性；如果系统的阻挫键类型处于无序状态，则$\psi_H$的模长将趋于零，序参量分布不再具有方向性。

综上，本文建立的这两类序参量是对正方晶格序参量的探索性推广：从多边形阻挫单元出发，找到阻挫键数量最少的情况，并分别将这些情况推广至整个蜂窝系统。随后再找出系统属于基态的情况。当系统处于对于这些基态时，序参量模长应达到最大；当系统脱离这些基态，进入无序的自旋分布时，序参量模长应趋于零。基于这样的考虑，构造了上述的（1-9）和（1-11）的形式。

## 1.3 小结

本文将设置低温条件，引入横向磁场产生量子相变，测量六角蜂窝晶格的序参量变化。在横磁场较低的情况下，系统主要由伊辛模型的磁耦合能决定，此时序参量的模长应该较大，且具有方向性。增大横向磁场会加剧量子涨落，系统表现得更无序[11]，序参量的模长应当变小、方位性逐渐消失；同时选定不同尺寸的系统进行对比[12]。

量子计算机对这种复杂量子相变的模拟有着很大潜力，但如今这种新型计算机并未普及，相关性能有待考察，需要经典计算机的结果作对照。因此，本文在经典计算机上采用量子蒙特卡洛法来模拟复杂的量子相变，以期为量子计算机的结果提供参考。

# 第 2 章 模型与原理

## 2.1 六角蜂窝晶格模型

### 2.1.1 横场伊辛模型

本文的几何阻挫下的二维六角蜂窝晶格，采用横向磁场下的伊辛模型描述该体系的哈密顿量。把实际的二维六角蜂窝材料的磁单元抽象为格点，格点具有不同自旋态，它们之间存在自旋磁矩带来的磁耦合能。晶格整体的能量来源于磁耦合能以及在格点在外磁场中的势能。采用简化模型，用z方向的耦合能代表格点间势能，把 N 格点系统的哈密顿量 H 写为下式：

$$H = \sum_{<i,j>} J_{ij}\, \sigma_i^z \sigma_j^z + g \sum_i \sigma_i^x \tag{2-1}$$

其中，i、j 是格点编号，取 1 到N。<i, j>表示只考虑近邻格点间的耦合能，$J_{ij}$是磁耦合系数，g是横向磁场大小。$\sigma_i^z$，$\sigma_i^x$ 分别代表z以及x方向的自旋泡利矩阵。系统以z方向的自旋本征态：$|\alpha\rangle = |\sigma_1, \sigma_2, \sigma_3, ..., \sigma_N\rangle$作为基，用|1⟩标记格点的上自旋态，|−1⟩标记格点的下自旋态，格点z方向的自旋值相应记为1、−1。例如，所有格点都是上自旋态，则$|\alpha\rangle=|1, 1, ..., 1\rangle$，格点自旋值都为 1。在磁场g较小时，系统倾向处于伊辛耦合项描述的基态，格点自旋分布主要由伊辛耦合项描述，具有特定的次序；当磁场g较大时，格点自旋分布趋于随机取±1，系统处于无序的状态。

### 2.1.2 结构配置

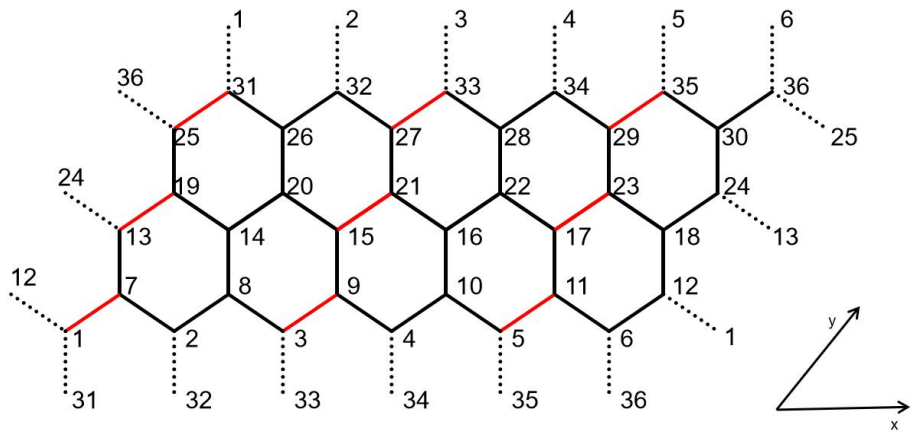

图 8 尺寸为 5×2 的六角蜂窝晶格

上图 8 展示了一个六边形单元数量为 5×2（Lx×Ly）的系统，一共有 36 个格点。黑

色键表示铁磁键，红色线表示反铁磁键，虚线表示周期边界条件。为了同时满足上一章节序参量的格点编号形式，本文研究的尺寸为：Lx = 5 + 6m，Ly = 2 + 3n，m和n是非负整数。既定尺寸下，格点数nn = 2(Lx + 1)(Ly + 1)，伊辛耦合键数nb = 3(Lx + 1)(Ly + 1)。

## 2.2 量子蒙特卡洛方法

关于六角蜂窝晶格序参量及其他参数的测量，可以借助量子蒙特卡洛方法完成。本节主要介绍了随机级数展开法的原理、马尔科夫平衡链以及程序思路。

### 2.2.1 随机级数展开法

$$\rho_a \propto e^{-\beta H_a} \tag{2-2}$$

在正则系综下考察哈密顿量为H的六角蜂窝结构，系统的能量本征态的应该满足热学分布(2-2)。上式$\beta = 1/k_b T$，代表逆温度。为了获取物理量$\langle u \rangle$，一般要找到系统的能量本征态a，对本征态a的物理量$u_a$做关于权重$\rho_a$的加权平均。然而问题在于，经典计算机以±1表示系统的自旋本征态$|\alpha\rangle$，并不是横场伊辛模型的能量本征态，不能直接利用权重$\rho_a$进行加权；此外，阻挫六角蜂窝晶格属于及其复杂的系统，很难找到能量本征态的解析形式——现如今只有少量伊辛模型可以解析求解。对于这种情况，可以引入随机级数展开法，将待测物理量（某u）的热学期望值改写成能用本征态$|\alpha\rangle$表示的权重求和的形式。热学期望值$\langle u \rangle$如下式(2-3)，其中Z是配分函数，Tr是对矩阵算符求迹。

$$\langle u \rangle = \frac{1}{Z} \text{Tr}(\hat{u} e^{-\beta \hat{H}}) \tag{2-3}$$

如(2-4)所示，依次为空算符、磁场算符、常数算符以及伊辛算符类型的格点算符，其中v代表算符类型，k代表格点i编号以及近邻格点对的i, j编号。将H分解为格点算符$H_{v,k}$：$H = -\sum_i (H_{h,i} + H_{c,i}) - \sum_{\langle i,j \rangle} H_{Is,i,j}$，H只包含后面三种算符[13][14]。

$$H_{0,0} = 1 \tag{2-4 a}$$

$$H_{h,i} = g(\sigma_i^+ + \sigma_i^-), \quad i > 0 \tag{2-4 b}$$

$$H_{c,i} = g, \quad i > 0 \tag{2-4 c}$$

$$H_{Is,i,j} = |J_{i,j}| - J_{i,j} \sigma_i^z \sigma_j^z, \quad i,j > 0 \tag{2-4 d}$$

将展开的哈密顿量H带入配分函数(2-3)，用一组完备的基$|\alpha_0\rangle$，把$\langle u \rangle$的级数展开写为下式，

$$\langle u \rangle = \sum_{\alpha_0} \sum_{n=0}^{\infty} \sum_{C_n} \frac{\beta^n}{Z \cdot n!} \langle \alpha_0 | \hat{u} \prod_{r=1}^{n} \hat{H}_{v,k_r} | \alpha_0 \rangle = \sum_{\alpha_0} \sum_{n=0}^{\infty} \sum_{C_n} W \times U \quad (2\text{-}5)$$

$$W = W(\alpha_0, n, C_n) = \frac{\beta^n}{Z \cdot n!} \langle \alpha_0 | \prod_{r=1}^{n} \hat{H}_{v,k_r} | \alpha_0 \rangle \quad (2\text{-}6)$$

$$U = U(\alpha_0, n, C_n) = \langle \alpha_0 | \hat{u} \prod_{r=1}^{n} \hat{H}_{v,k_r} | \alpha_0 \rangle \Big/ \langle \alpha_0 | \prod_{r=1}^{n} \hat{H}_{v,k_r} | \alpha_0 \rangle \quad (2\text{-}7)$$

上式(2-5)只考虑不为零的$W(\alpha_0, n, C_L)$的求和，函数U也只对这些$(\alpha_0, n, C_L)$定义。$\prod_{r=1}^{n} \hat{H}_{v,k_r}$是按一定顺序$C_n$作用的格点算符的序列；W是归一化的，具有权重的意义，称为权重函数。考虑到计算机模拟的局限，为了表示出$\langle u \rangle$，要在L处将级数截断，取前面的部分多项式代替整个级数。L和系统的格点数量、温度有关，至少为βN的量级。截断后的$\langle u \rangle$展开为：

$$\langle u \rangle = \sum_{\alpha_0} \sum_{n=0}^{L} \sum_{C_n} \frac{\beta^n}{Z \cdot n!} \langle \alpha_0 | \hat{u} \prod_{r=1}^{n} \hat{H}_{v,k_r} | \alpha_0 \rangle = \sum_{\alpha_0} \sum_{C_L} \frac{\beta^n (L-n)!}{Z \cdot L!} \times \langle \alpha_0 | \hat{u} \prod_{r=1}^{L} \hat{H}_{v,k_r} | \alpha_0 \rangle$$
$$(2\text{-}8)$$

上式引入了(L-n)个空算符$H_{0,0}$，把它们插入$C_n$代表的算符序列$\prod_{r=1}^{L} \hat{H}_{v,k_r}$，将后者扩充为L个算符的序列$C_L$。对$C_L$序列的所有情况求和，同时消去倍数$L!/n!(L-n)!$的影响；这里的n是$n(C_L)$简写，即非空算符的数量。结合(2-5)改写(2-8)，插入(L-1)个单位矩阵：$I = \sum_{\alpha_0} | \alpha_p \rangle \langle \alpha_p |$并化简，

$$\langle u \rangle = \sum_{\alpha_0} \sum_{C_L} \frac{1}{Z} \cdot \frac{\beta^n (L-n)!}{L!} \langle \alpha | \prod_{r} \hat{H}_{v,k_r} | \alpha \rangle \times \frac{\langle \alpha_0 | \hat{u} \prod_{r} \hat{H}_{v,k_r} | \alpha_0 \rangle}{\langle \alpha_0 | \prod_{r} \hat{H}_{v,k_r} | \alpha_0 \rangle}$$

$$= \sum_{\alpha_0} \sum_{C_L} \frac{1}{Z} \cdot \frac{\beta^n (L-n)!}{L!} \sum_{\{\alpha_p\}} \langle \alpha_L | \hat{H}_{v,k_L} | \alpha_{L-1} \rangle \cdots \langle \alpha_1 | \hat{H}_{v,k_1} | \alpha_0 \rangle \times \frac{\langle \alpha_0 | \hat{u} \prod_{r} \hat{H}_{v,k_r} | \alpha_0 \rangle}{\langle \alpha_0 | \prod_{r} \hat{H}_{v,k_r} | \alpha_0 \rangle}$$

$$= \sum_{\alpha_0} \sum_{C_L} \frac{1}{Z} \cdot \frac{\beta^n (L-n)!}{L!} \sum_{\{\alpha_p\}} \prod_{p=1}^{L} \langle \alpha_p | \hat{H}_{v,k_p} | \alpha_{p-1} \rangle \times \frac{\langle \alpha_0 | \hat{u} \prod_{r=1}^{L} \hat{H}_{v,k_r} | \alpha_0 \rangle}{\langle \alpha_0 | \prod_{r=1}^{L} \hat{H}_{v,k_r} | \alpha_0 \rangle}$$

$$= \sum_{\alpha_0} \sum_{C_L} \frac{1}{Z} \cdot \frac{\beta^n (L-n)!}{L!} \sum_{\{\alpha_p\}} \prod_{p=1}^{L} \langle \alpha_p | \hat{H}_{v,k_p} | \alpha_{p-1} \rangle \times \frac{\langle \alpha_0 | \hat{u} \prod_{r=1}^{L} \hat{H}_{v,k_r} | \alpha_0 \rangle}{\langle \alpha_0 | \prod_{r=1}^{L} \hat{H}_{v,k_r} | \alpha_0 \rangle}$$

$$= \sum_{\{\alpha\}} \sum_{C_L} \frac{1}{Z} \cdot \frac{\beta^n (L-n)!}{L!} \prod_{p=1}^{L} \langle \alpha_p | \hat{H}_{v,k_p} | \alpha_{p-1} \rangle \times \frac{\langle \alpha_0 | \hat{u} \prod_{r=1}^{L} \hat{H}_{v,k_r} | \alpha_0 \rangle}{\langle \alpha_0 | \prod_{r=1}^{L} \hat{H}_{v,k_r} | \alpha_0 \rangle}$$

$$= \sum_{\{\alpha\}} \sum_{C_L} W(\alpha, C_L) \times U(\alpha, C_L) \quad (2\text{-}9)$$

如上(2-9)，$\{\alpha_p\}$代表中间插入的长度为(L-1)的本征态$|\alpha_p\rangle$序列，且$|\alpha_0\rangle = |\alpha_L\rangle$；在$\{\alpha_p\}$的基础上合并$|\alpha_0\rangle$，形成$\{\alpha\}$，代表L个本征态$|\alpha\rangle$的序列：$|\alpha_0\rangle$、$|\alpha_1\rangle$……$|\alpha_L\rangle$。

对任意的$\langle \alpha_p | \hat{H}_{v,k_p} | \alpha_{p-1} \rangle$，本征态$|\alpha_{p-1}\rangle$经过格子算符$\hat{H}_{v,k_p}$作用后依旧是某种本征态$|\alpha\rangle$，由于正交性，如果后面的态$|\alpha_p\rangle$与其不同，则$\langle \alpha_p | \hat{H}_{v,k_p} | \alpha_{p-1} \rangle$为零，相应的$W(\alpha, C_L)$权重也为零。和(2-5)类似，(2-9)只考虑非零项的求和。对于某个集合α，想要得到不为零的权重，只能满足$|\alpha_p\rangle = \hat{H}_{v,k,p} | \alpha_{p-1} \rangle$。因此，基于$|\alpha_0\rangle$和$\hat{H}_{v,k,p}$序列即可得到满足要求的集合α：$|\alpha_0\rangle$、$|\alpha_1\rangle$……$|\alpha_L\rangle$。对集合中的$|\alpha_p\rangle$重新排列，只要符合循环方向$|\alpha_0\rangle \rightarrow |\alpha_1\rangle \rightarrow ... |\alpha_L\rangle \rightarrow |\alpha_0\rangle \rightarrow ...$，都可以成为新的集合。上式(2-9)就可以看作对所有$|\alpha_0\rangle$、

以及不同的$C_L$算符序列进行的求和。提取W、U：

$$W(\alpha_0, C_L) = \frac{1}{Z} \cdot \frac{\beta^n (L-n)!}{L!} \prod_{p=1}^{L} \langle \alpha_p | \hat{H}_{v,k_p} | \alpha_{p-1} \rangle \quad (2\text{-}10)$$

$$U(\alpha_0, C_L) = \langle \alpha_0 | \hat{u} \prod_{r=1}^{L} \hat{H}_{v,k_r} | \alpha_0 \rangle \Big/ \langle \alpha_0 | \prod_{r=1}^{L} \hat{H}_{v,k_r} | \alpha_0 \rangle \quad (2\text{-}11)$$

本文测量的初级序参量$m_H$以及次级序参量$\psi_H$，其算符是关于自旋本征态$|\alpha\rangle$的对角算符，代入$U(\alpha, C_L)$后得到下式（2-12）；$h_r$是格点算符$\hat{H}_{v,k_r}$作用于传播态$|\alpha_p\rangle$产生的系数。实际的蒙特卡洛模拟采用$U_{ave}$，它表明对所有满足$|\alpha_p\rangle$的循环方向结果取平均：(2-13)，

$$U(\alpha_0, C_L) = \langle \alpha_0 | \hat{u} \prod_{r=1}^{L} h_r | \alpha_0 \rangle \Big/ \langle \alpha_0 | \prod_{r=1}^{L} h_r | \alpha_0 \rangle = \langle \alpha_0 | \hat{u} | \alpha_0 \rangle \quad (2\text{-}12)$$

$$U_{ave} = \frac{1}{L} \sum_{p=0}^{L-1} \langle \alpha_p | \hat{u} | \alpha_p \rangle \quad (2\text{-}13)$$

可见，原本对热学期望值$\langle u \rangle$的测量，是找出能量本征态a后按照$\rho_a \propto e^{-\beta H_a}$的分布进行统计平均。但由于计算机模拟的是z方向的自旋本征态$|\alpha\rangle$，并且很难求出能量本征态a的解析解，无法直接用$|\alpha\rangle$表示出a，也就无法使用加权系数$\rho_a$。这里借鉴随机级数展开法，改写了$\langle u \rangle$的形式：$\sum_{\alpha_0, C_L} W(\alpha_0, C_L) \times U(\alpha_0, C_L)$，将期望值分解为与本征态$|\alpha\rangle$的函数W、U，使$\langle u \rangle$的自变量从能量本征态转变为自旋本征态$|\alpha\rangle$。

为了获取$\langle u \rangle$，要按照权重$W(\alpha_0, C_L)$对函数$U(\alpha_0, C_L)$取平均。由于六角蜂窝晶格系统的本征态数量极其庞大，再加上传播态序列的展开，待测样本数远超计算机的计算范围，因而不能采用遍历的方法求值。像这样的问题，可以凭借蒙特卡洛方法解决：先随机抽样获取大量样本，再让样本群体满足特定的权重分布，最后测量群体的统计值。蒙特卡洛方法的关键点在于使随机样本符合权重分布，要做到这一点这可以构建马尔科夫链。

## 2.2.2 构建马尔科夫链

马尔科夫链是一种状态序列，序列中的某个状态仅与上一个状态有关。设初始时刻的随机样本分布为$S_0 = (q_1, q_2, ..., q_n)^T$，总样本数为n，样本i出现的概率是$q_i$，满足归一化。样本分布S经过跃迁矩阵$P = \{P_{ij}\}$作用后到达下一个状态S'。矩阵元$P_{ij} = P(j|i)$，表示下一时刻样本 j 跃迁到样本 i 的概率。如下(2-14)式描述了一个马尔科夫链。跃迁矩阵P应该满足(2-15)式，

$$S_0 \rightarrow P^1 S_0 \rightarrow P^2 S_0 \rightarrow \cdots\cdots \rightarrow P^k S_0 \quad （2\text{-}14）$$

$$\lim_{k \to \infty} (P^k)_{ij} = W_i \quad （2\text{-}15）$$

此时，不论初始$S_0$是何种样本分布，经过$P^k$次跃迁后都会得到要求的平衡分布：$S_f = (W_1, W_2, ..., W_n)^T$。因此，只要构建出满足上式的跃迁矩阵，即可使得初始样本分布在多次

跃迁后达到权重W的分布。

利用细致平衡准则：$W_i P_{ji} = W_j P_{ij}$，即可得到满足条件（2-15）的矩阵元。跃迁概率 $P_{ij}$ 可写为选择概率 $A_{ij}$ 与接受概率 $B_{ij}$ 的乘积：$W_i \cdot A_{ij} \cdot B_{ji} = W_j \cdot A_{ij} \cdot B_{ij}$，其中 $A_{ij}$、$B_{ij}$ 的定义与 $P_{ij}$ 类同。细致平衡准则的含义是，当系统处于平衡时，从任意一个样本 i 跃迁到任意样本 j 的几率流与从 j 跃迁到 i 的几率流是相等的。几率流便取决于位形的权重以及跃迁过程的选择概率和接受概率。根据Metropolis算法准则的接受概率为，

$$B_{ij} = \min(1, \frac{W_i \cdot A_{ji}}{W_j \cdot A_{ij}})\quad (2\text{-}16)$$

实际模拟中，首先生成一个随机样本$a(0) = i$，再依据选择概率 $A_{ji}$ 随机生成下一个样本$a(1) = j$，然后利用接受概率$B_{ji}$判断是否接受新样本j。像这样进行大量的k次迭代后，可以认为样本的分布达到了平衡。随后继续迭代多次得到n个样本，这些样本即满足平衡分布：$a(k), a(k+1)......, a(k+n)$ 对应于W权重的分布。

本文的样本是六角蜂窝晶格本征态$|\alpha\rangle$的传播序列$(\alpha_0, C_L)$，权重为$W(\alpha_0, C_L)$。哈密顿量H拆分为的格点算符可以保证权重W非负，并且在程序中排除了权重W为零的样本$(\alpha_0, C_L)$。因此，程序关注的样本$(\alpha_0, C_L)$的权重$W > 0$，从任意一个$(\alpha_0, C_L)$出发，都能通过非零的概率跃迁到任意的另一个$(\alpha_0', C_L')$，从而实现遍历性。样本$(\alpha_0, C_L)$之间的跃迁是在不同算符之间进行的单个替换，分为空算符$H_{0,0}$与伊辛算符$H_{Is,i,j}$、空算符$H_{0,0}$与常数算符$H_{c,i}$、磁场算符$H_{h,i}$与常数算符$H_{c,i}$之间的替换。其中，$H_{0,0}$换为$H_{Is,i,j}$和$H_{c,i}$的选择概率分别是：$P_{in,IS}$、$P_{in,c}$，且$P_{in,IS} + P_{in,c} = 1$[15]；而$H_{Is,i,j}$和$H_{c,i}$换为$H_{0,0}$的选择概率设为1。$H_{h,i}$、$H_{c,i}$替换前后样本$(\alpha_0, C_L)$的权重W相等，于是把它们之间的选择概率都设为1。将选择概率带入(2-16)即可得到相应的接受概率。

## 2.3 算法介绍

基于随机级数展开法[14]，在fortran90上实现的程序大致如下。

**步骤1**：数据初始化。

定义主程序变量：伊辛步isteps、统计次数nbins、单次统计的蒙特卡洛步数mstep、自旋矩阵spin、算符序列opstring、顶点列表vertexlist等，外加磁场hh、晶格尺寸Lx、Ly等参数通过调用文件获取。随后构建本文的阻挫六角蜂窝晶格，设计生成任意L×L的晶格程序，并绘图示意。初始化spin(:),格点自旋随机设为±1。算符序列长度mm初始化为20，全为空算符。

**步骤2**：建立分布平衡，在循环i=1:isteps中，依次执行下述四个步骤。伪代码如下，

```
for i=1:isteps
    do update          %更新算符序列
    do linkvertex      %顶点链接
```

```
    do clusterupdate    %自旋团簇翻转
    do adjustlayer      %序列长度修正
end                     %达到平衡分布
```

(2.1)更新算符序列：沿着传播态传播态序列方向逐层更新算符序列$C_L$。如果本层是空算符，则生成[0,1]的均匀随机数r=rand()，如果$r < P_{in,c}$则选择常数算符，否则选伊辛算符。随后继续生成均匀随机数r，若$r < P_{accept,c}$则接受跃迁，否则保持不变。如果本层是伊辛（或常数）算符，生成均匀随机数r，如果$r < P_{accept,c\_0}(P_{accept,IS\_0})$则替换为空算符，否则不进行替换。如果本层是磁场算符，则翻转本层中i格点的自旋，其跃迁在下述(2.3)进行。

(2.2)格点链接：遍历三维的传播态序列层，根据伊辛、常数、磁场算符的特性，将具有相同自旋状态且毗邻的格点链接起来，用vertexlist列表记录格点之间的链接关系。

(2.3)自旋团簇翻转：根据vertexlist列表把相互链接的格点记为一个团簇。对于每个团簇，生成随机数r=rand()，如果$r < 1/2$则标记为翻转，否则标记为不翻转。在标记翻转的过程中执行了磁场和常数算符的跃迁，且跃迁数量是偶数，保证了传播态序列$(\alpha_0, C_L)$的周期边界条件。传播态序列由初始层和既定的算符序列决定，初始层$\alpha_0$的自旋翻转即可实现整个传播态序列的团簇翻转。仅凭马尔科夫过程改变传播态序列的效率显得不足，所以依靠团簇更新对三维传播态序列层的自旋位形进行变化，让序列$(\alpha_0, C_L)$不局限于某种分布类型。

(2.4)序列长度修正：对于非空算符即将填满传播态序列的情况，应该适当增加算符序列长度$C_L$的长度$L_p$，保证非空算符数量$n_h$和$L_p$之间存在一定差异。这类情况预示着$(\alpha_0, C_L)$对于大$n_h$有着较大的权重，只有增加$L_p$才能保证之后的跃迁能把这类较大权重的情况容纳进来。设置增加条件为：$L_{new,p} = 10/9\, n_h + (L_x \cdot L_y)^2/100$。如果$L_{new,p} > L_p$则增加$L_p$，否则不增加。

**步骤3**：将平衡后的算符序列长度$L_p$、非空算符数$n_h$写入文件保存。

**步骤4**：结果统计。和步骤2相比，不再进行算符序列长度的更新。一共进行nbins次测量，每次测量都会得到一组待测物理量⟨u⟩，每组⟨u⟩是mstep次的蒙特卡洛统计结果。伪代码如下所示：

```
for i=1:nbins               %统计次数
    for j=1:mstep           %蒙卡步数
        do update           %更新算符序列
        do linkvertex       %顶点链接
        do clusterupdate    %自旋团簇翻转
        do measure          %测量结果
    end
    do write_results        %写入结果
```

end

**步骤 5**：释放数组，清空内存。

# 第 3 章 结果测量与分析

本文关注的是低温下的量子相变，逆温度 β = 1/T = 3.3，不断改变横向磁场值 g 以及系统尺寸 Lx×Ly 得到测量结果。格点间伊辛耦合强度：$J_{ij}$=±J，J 取 1，则下文用 g 代表 g/J。在满足周期边界条件的同时完成序参量的编号，则六边形阵列最小为 5 × 2（Lx，Ly），其格点数是 36。将这样的尺寸记为 L×L，L=1。本文主要分为两类模拟测量如下：

（1）改变磁场值：在六角蜂窝晶格尺寸为 36 个格点（L=1）的条件下，设置一系列横向磁场值：0、0.1、0.2、0.3、0.4、0.5、0.6、0.7、0.8、0.9、1.0，观察序参量分布情况和模长的变化。

（2）对比不同尺寸：设置一系列磁场值：0、0.2、0.4、0.6、0.8、1.0，在尺寸为 144 个格点（11×5 六边形，L=2）的条件下、尺寸为 324 个格点（17×8 六边形，L=3）的条件下进行测量，对比能量密度和序参量的情况。

## 3.1 横向磁场的影响

### 3.1.1 序参量的分布图像

以 g=0、0.1、0.2、0.6 和 2.0 为例，$m_H$ 和 $\psi_H$ 的分布图像如下所示，

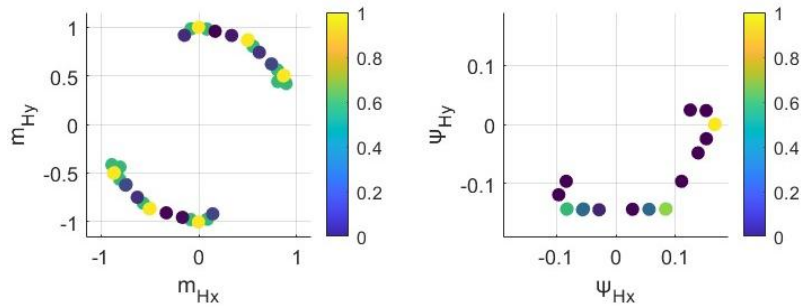

图 9.1　g=0

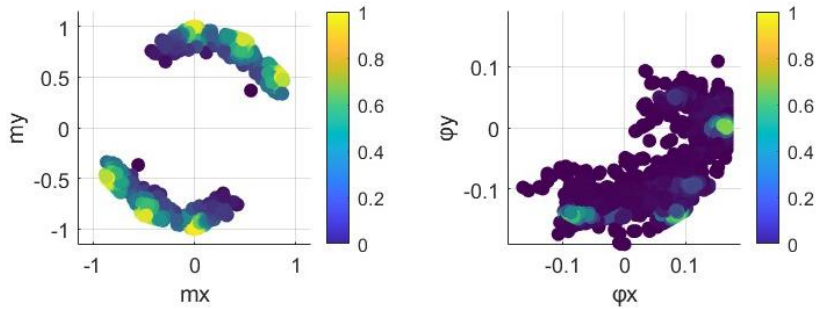

图 9.2　g=0.1

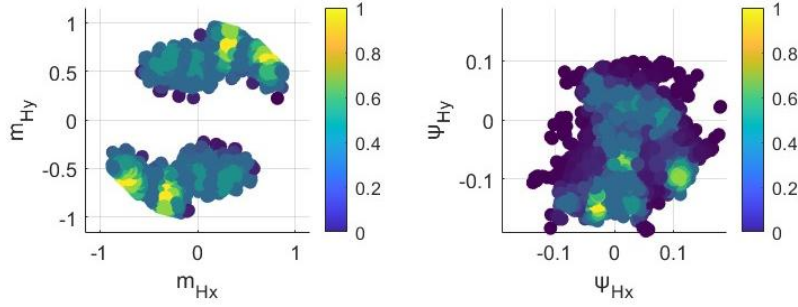

图 9.3    g=0.2

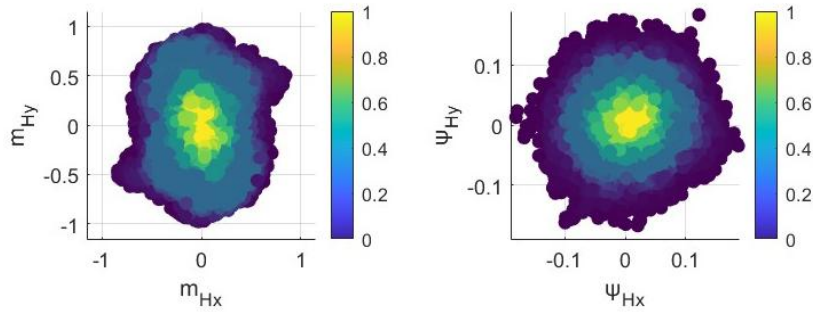

图 9.4    g=0.6

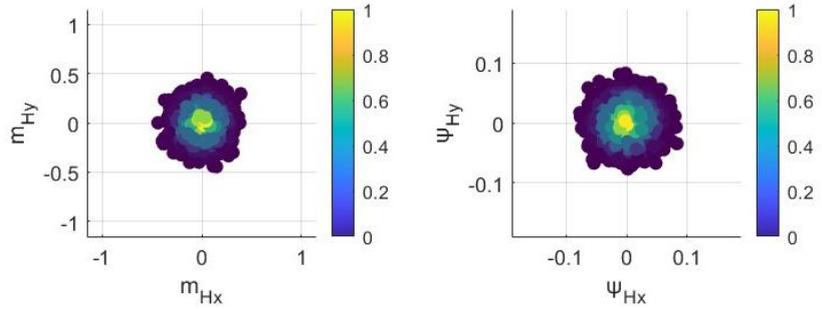

图 9.5    g=2.0

为了系统局限于拓扑缺陷，程序设置的初始自旋位形是上文提及的六种基态。随后得到序参量在马尔科夫链下的平衡分布。由上图可见，在较低的磁场条件下，初级序参量$m_H$呈现六个集中分布的方位：$1/6\pi$、$2/6\pi$、$3/6\pi$、$7/6\pi$、$8/6\pi$和$9/6\pi$；次级序参量$\psi_H$呈现三个集中分布的方位：$4/3\pi$、$5/3\pi$和$6/3\pi$，与预期结果相符合。随着磁场的增加，序参量逐渐收缩，在较大磁场下变为集中于原点的圆斑，符合磁场增加系统格点z方向自旋随机性增加的特点。

上图9.3代表了序参量分布的转变过程。以初级序参量$m_H$为例，$m_H$的分布并非同时从六个方向往中心收缩，进而得到一个更小的圆弧；它呈现了特殊的收缩方式，如下图9.6

所示：红色圈可能代表能量很高的状态，在适中的磁场条件下，系统很难到达该区域，无法经过该区域往原点方向收缩；红色箭头代表主要的收缩方向，系统可以沿着该方向到达能量比较高的区域。次级序参量$\psi_H$的结果也是类似的。红色箭头是通过观察序参量平衡分布下的数据得到的大致结论。

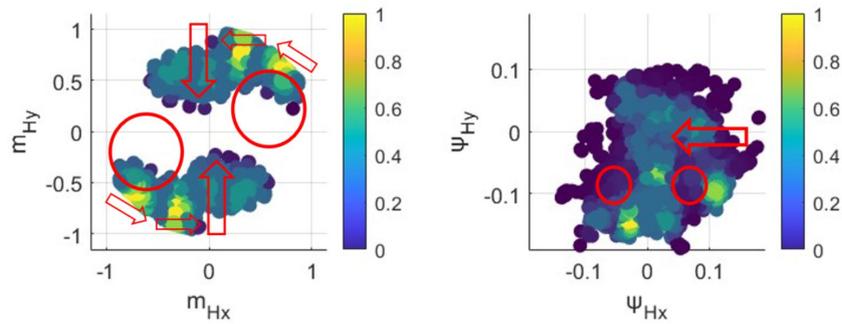

图 9.6  g=0.2 的转变过程

### 3.1.2 序参量模长随磁场的变化

如下图所示，模长随横向磁场的增加而减小。在低磁场条件下，$|m_H| \cong 1$ 且 $|\psi_H| \cong 1/6$，符合理论预期。在磁场g约为0.1附近完成模长的突变，随后趋于零值。在g=0.4 到 0.5 之间存在平台期，模长变化幅度较小，可能因为系统在该磁场范围下难以进一步跃迁到能量更高的位形上，被相对局限于该磁场范围下所处的位形集合。

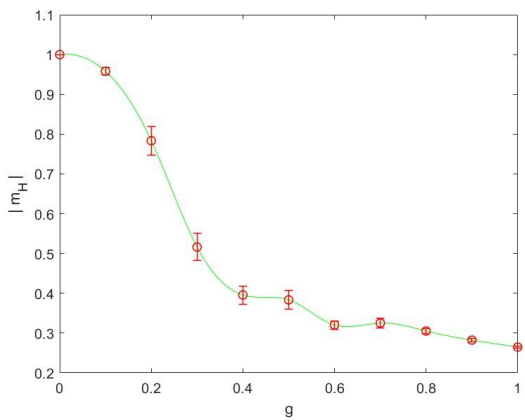 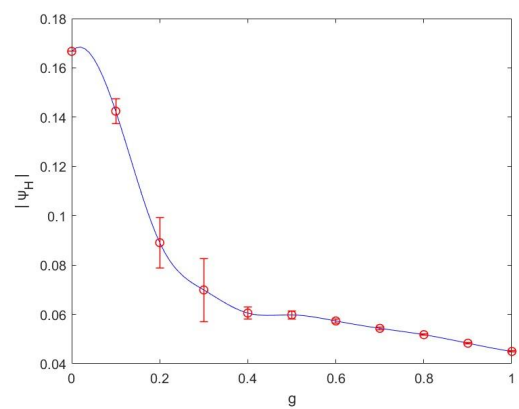

图 10.1  初级序参量模长$|m_H|$　　　　　图 10.2  次级序参量模长$|\psi_H|$

## 3.2 不同尺寸结果

### 3.2.1 能量密度的变化

$$e = \frac{\langle E \rangle}{nn} = -\frac{\langle n(C_L) \rangle}{\beta \cdot nn} + h + \sum_{\langle i,j \rangle} \frac{|J_{ij}|}{nn} \tag{3-1}$$

基于上一章的随机级数展开法得到式(3-1)[14]，可以得到能量密度的结果如下图 11 所示。可见，随着横磁场的增加晶格整体的能量密度减小，说明系统的格点磁矩逐渐趋于磁场x方向的排列，在外加横向磁场作用下，具有的势能逐步降低，处于外加磁场下的稳定状态。不同尺寸大小的系统得到的能量密度一致，三种颜色的测量曲线重叠在了一起。

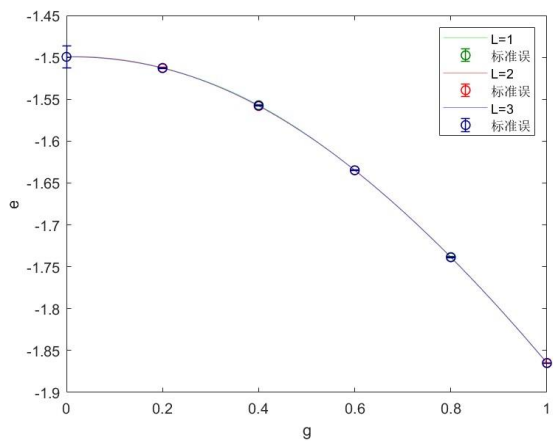

图 11 能量密度随磁场的变化

### 3.2.2 序参量分布的变化

图 12 以h=0 的情况为例。尺寸较小时系统的序参量分布比较发散，展现出了圆弧的样式，$m_H$在预期的六个方位上分布较集中，$\psi_H$则对应地分布于三个预期的方位。随着尺寸的增加，序参量越发集中于特定的方位上。本文设置的平衡步长和测量次数均与系统的尺寸成正比关系。

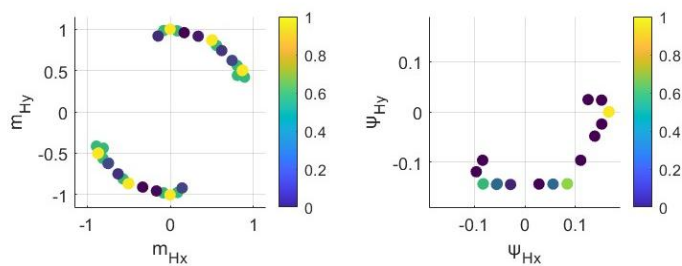

图 12.1　L=1（36 格点）

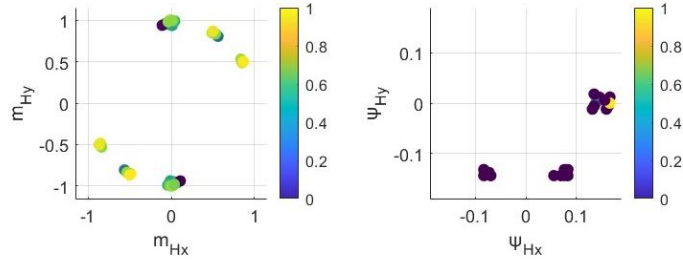

图 12.2　L=2（144 格点）

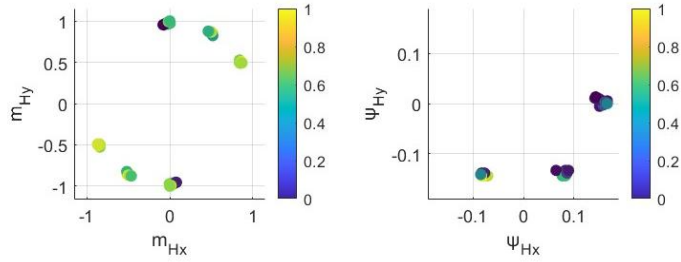

图 12.3　L=3（324 格点）

### 3.2.3　序参量模长的对比

下图 13 展示了 β = 3.3 时，系统自初始位形经历较大的平衡步数后获得的结果，在这样的序参量编号方式下，序参量随着尺寸的增加在横向磁场 g 较小时（例如 g=0.2）迅速趋于零，并没有在关注的温度和磁场条件下出现模长不为零的平台期，因此暂时无法确定相变点的位置。一种可能的解释为：加入磁场条件后，系统从本文选定的初始基态自旋位形出发，最终到达的平衡分布包含大量其他位形。图 14 展示了其他位形和基态位形的可能关系，本文的情况如图（a）所示，系统一旦跳出本文选定的基态位形，进入其他位形，再次返回基态位形的可能性很低，而其他位形对应的序参量模长很小，因此模长随磁场的增加很快的趋于零。此外本文仅从选定的基态位形出发，并未考察其他的基态情况。最终落

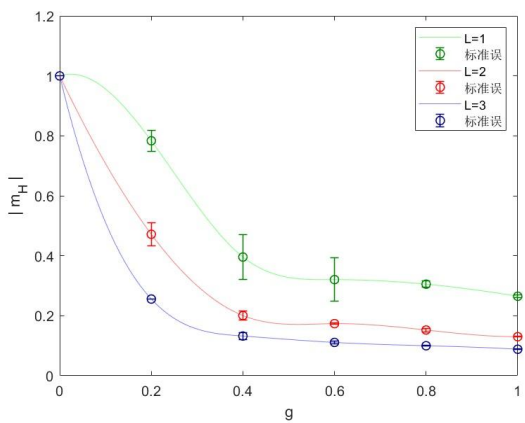
图 13.1　|$m_H$|的对比图

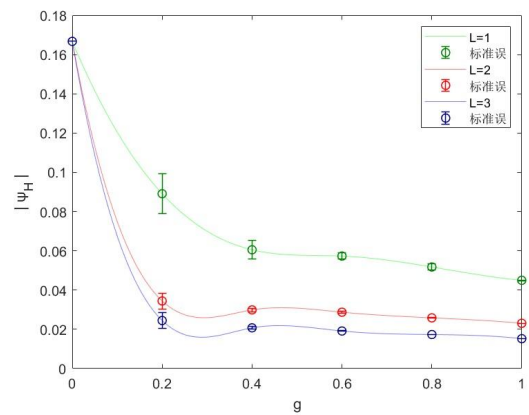
图 13.2　|$\psi_H$|的对比图

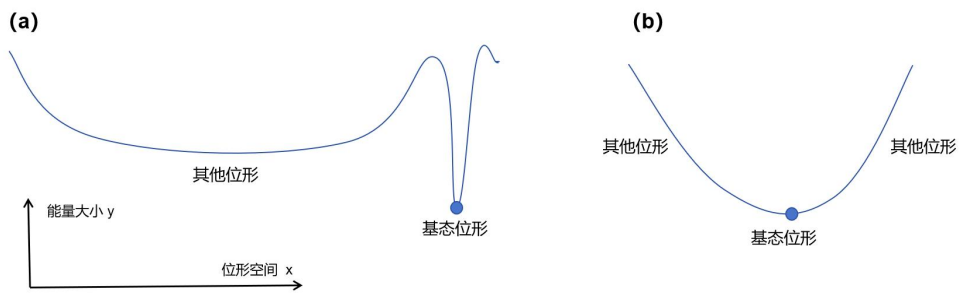

图 14 位形-能量分布图

（位形之间的横轴距离可以表示位形相似性，距离越近相似性越高）

入的平衡分布可能包含大量的其他的基态，而这些基态对应的序参量模长很小，因此统计到的期望值很小。

相比起来，以往正方晶格的研究可能类似于图 14 的情况（b），系统从选定的基态位形出发，进入其他位形后再次返回基态位形的可能性较大（局域最小值不明显），最终呈现的平衡分布集中于基态位形，其序参量模长值在h非零的情况下也不会快速趋于零。

# 第 4 章 总结与展望

## 4.1 总结

本文建立了一种几何阻挫的六角蜂窝晶格,在横场伊辛模型下探究了它的量子相变性质。基于阻挫三角晶格、阻挫正方晶格序参量的形式,设置了描述六角蜂窝晶格相变的初级序参量$m_H$和次级序参量$\psi_H$,以便分析六边形单元的格点自旋分布、受挫键类型。之后介绍了随级数展开法,经过数值模拟测量可以得到如下结论:

(1)能量密度:不同尺寸下的能量密度曲线图重叠在了一起,随着横向磁场的增加,能量密度逐渐减小,符合理论预期。

(2)序参量:在 $\beta = 3.3$,$h = 0$ 的条件下,系统的初级序参量$m_H$分布呈现了六种不同的方位,次级序参量$\psi_H$则呈现了三种对应的方位,说明此时系统局限于选定的基态自旋位形。随着磁场的增加,序参量的模长迅速趋于零值。这意味着系统的位形-能量分布可能如上图 14(a)所示,一旦跳出选定基态位形,将难以返回到原有位形,且平衡分布下的大多数位形对应的序参量模长非常小,因此磁场增加后模长快速趋近零,在本文选定的磁场条件下暂时无法确定相变点。并且其他位形中可能包含大量的基态,也要排除这些基态的影响。

(3)位形-能量分布的区别:通过六蜂窝角晶格和正方晶格的对比,可能说明两种系统的位形-能量分布有较大差异。平衡分布下,六角蜂窝晶格的多数位形并未处于选定的基态位形附近,与选定的基态位形相似度低;而平衡分布下的正方晶格的多数位形处于八种基态位形的附近,与基态位形的相似度高。

## 4.2 展望

本文的研究内容可以从以下几个方向作完善和拓展:(1)复现正方晶格的相关研究,考察其序参量随磁场增加的分布情况,并与本文的结果做对比。(2)极低温度、极小磁场:在极低温度和极低磁场的条件下,考察系统序参量模长的变化情况。(3)结合退火算法进行:正方晶格选定的基态位形和其他位形的关系可能如上图 14(b)所示,因此即使在非零的磁场下,平衡分布带来的序参量模长也不会迅速趋于减小,所以可以利用退火算法详细考察正方晶格的基态位形以及六角蜂窝晶格的基态位形,并结合基态对六角晶格的序参量做出改进。

# 参考文献

# 致 谢

  衷心感谢邵慧老师和刀流云师兄对我悉心的学术指导。邵老师在我学习量子退火原理、量子相变的过程中循循善诱，带领我接触到了量子计算这个迷人的、富有活力的新兴领域，给了我学术上极大的帮助以及精神上的鼓励。刀师兄在相关的算法问题上给了我很大的支持，他帮助我解决了编程中遇到的难题，悉心负责的教导让我受益匪浅，如沐甘霖。学为人师，行为世范，他们的教诲如春风化雨，让我有了很大的收获。

  同时，也要感谢家人朋友，他们在生活中给了我无微不至的关心和照顾，让我在学习前进的路上倍感欣慰。

  最后，还要感谢身边的同学，每当遇到学习上的困难，都会和他们谈论探究，相互学习，共同进步。

<div style="text-align:right">

陈自立

2024 年 5 月

</div>